\documentclass{article}

\usepackage{amsmath, amssymb} %,amsthm,txfonts
\usepackage{tensor}
\usepackage[small,nohug,heads=LaTeX]{diagrams}
\DeclareMathOperator*{\bigwp}{\text{\Large $\wp$}}
\newcommand{\bc}{\begin{C}}
\newcommand{\ec}{\end{C}}
\newcommand{\be}{\begin{equation}}
\newcommand{\ee}{\end{equation}}

\newcommand{\claim}{\begin{Cl}}
\newcommand{\eclaim}{\end{Cl}}
\newcommand{\nb}{\begin{Nb}}
\newcommand{\nbe}{\end{Nb}}
\newcommand{\bl}{\begin{LE}}
\newcommand{\el}{\end{LE}}

\newtheorem{Cl}{Claim}

\newcommand{\bd}{\begin{Def}}
\newcommand{\ed}{\end{Def}}
\newcommand{\bt}{\begin{Th}}
\newcommand{\et}{\end{Th}}

\newtheorem{Th}{Theorem}
\newtheorem{LE}{Lemma}
\newtheorem{C}{Corollary}
\newtheorem{Nb}{Note}
\newtheorem{Def}{Definition}

\def\Id{{\rm id}}
\def\mix{{\rm mix}}
\def\Mix{{\rm Mix}}
\def\ev{{\rm ev}}
\def\coev{{\rm coev}}
\def\Hid{{\rm Hid}}

\begin{document}
 \title{On partial traces and compactification of $*$-autonomous Mix-categories
}
\author{Sergey Slavnov
\\  National Research University Higher School of Economics
\\Moscow, Russia
\\ sslavnov@yandex.ru\\} \maketitle

\begin{abstract}
We study the question when a $*$-autonomous Mix-category has a representation as a $*$-autonomous Mix-subcategory of a compact one.
We define certain partial trace-like  operation on morphisms of a Mix-category, which we call a mixed trace, and show that any structure preserving embedding of a Mix-category into a compact one induces a mixed trace on the former. We also show that, conversely, if a Mix-category ${\bf K}$ has a mixed trace, then we can construct a compact category and structure preserving embedding of ${\bf K}$ into it, which induces the same mixed trace.

Finally, we find a specific condition expressed in terms of interaction of Mix- and coevaluation maps on a Mix-category ${\bf K}$, which is necessary and sufficient for a structure preserving embedding of ${\bf K}$ into a compact one to exist. When this condition is satisfied, we construct a ``free'' or ``minimal'' mixed trace on ${\bf K}$ directly from the Mix-category structure, which gives us also a ``free'' compactification of ${\bf K}$.
\end{abstract}

\section{Introduction}
$*$-Autonomous categories, monoidal categories with a particularly well-behaved duality, introduced by Barr \cite{Barr} are known in logic and computer science literature as models of linear logic, but, of course, they deserve interest on their own as well.

Compact (or compact closed) categories, are a particular subclass of $*$-autonomous categories, in which  duality preserves the monoidal structure. (The archetypical example is the category of finite-dimensional vector spaces, with monoidal structure given by the tensor product, and duality, the usual vector spaces duality.) They are studied a lot for  their own sake (i.e. without relation to general $*$-autonomous ones) and appear in very different contexts, such as categorical quantum mechanics, group representations, topology of manifolds and knot theory, theoretical computer science etc.

In logic and computer science literature, compact categories are particularly known for the {\it categorical trace},  a natural operation on morphisms, sending the map
$$f:A\otimes U\to B\otimes U$$
to its trace $$Tr(f):A\to B.$$
This operation satisfies a number of conditions and is
modeled after the usual linear operator trace in finite-dimensional vector spaces. In fact,  existence of categorical trace is a characteristic feature of compact categories \cite{JoyalStreetVerity}.
In theoretical computer science, trace, in particular,  is  used to model computational processes, feedback, cut-elimination etc.
  This  is usually discussed in the context of Girard's {\it Geometry of Interaction} \cite{Girard2} and, especially, its various subsequent ramifications such as \cite{Abramsky}, \cite{HS}.

  {\it Partial trace}, introduced in \cite{HS}, is a generalization of the ordinary (``total'') trace, which  satisfies basically the same properties, but is not necessarily defined for all morphisms. Primary motivation for this generalization comes, again, from Geometry of Interaction program. Partially traced categories are used to formulate categorical {\it Multi-object} Geometry of Interaction. As for relation to compact categories, it is proven in \cite{MalherbeScottSelinger, Bagnol}  that partially traced categories are precisely symmetric monoidal subcategories of compact (i.e. totally traced) ones: a symmetric monoidal category embeds into a compact one, if and only if it has a partial trace. The compact envelope of a partially traced category ${\bf K}$ can be constructed in a very transparent way \cite{Bagnol}: it has the same objects as ${\bf K}$, and its morphisms between $A$ and $B$ are ${\bf K}$-maps $A\otimes U\to B\otimes U$, quotiented by certain equivalence relation. The trace of a map $A\otimes U\to B\otimes U$ is then, modulo the above equivalence, the same map, but considered as a map $A\to B$.

  In this paper we study the question when a {\it $*$-autonomous Mix}-category has a structure preserving embedding  into a compact one. (Mix-categories \cite{Seely2} are a wide subclass of $*$-autonomous categories, more pedantically, of $*$-autonomous categories  with an extra structure. This class seems sufficiently wide to capture most of the ``general'' $*$-autonomous features, but is slightly easier to deal with.) 
  
  We see two (related) reasons why this question is interesting.

 First, many important $*$-autonomous categories have representations as subcategories of compact ones. In particular, in linear logic, a usual construction for building a non-degenerate model (compact categories, seen as models of linear logic, are degenerate) consists in some (often {\it ad hoc}) refinement of a given compact closed structure, which yields a new $*$-autonomous category, a subcategory of the initial compact one. The category of coherence spaces, which is the ``original'' model of linear logic, can be described in this way. Many other examples are considered in literature, see, say,  \cite{HylandSchalk}.
So, at least from the academic point of view, it is reasonable to ask if we can characterize $*$-autonomous categories of such a form.

Second, the above-mentioned Geometry of Interaction is closely tied to linear logic, and $*$-autonomous categories are models of linear logic, therefore it is desirable to understand them in one context. But Geometry of Interaction  is most often formulated in terms of categorical trace (total or partial), hence, eventually, an embedding into a compact category. It is natural to consider $*$-autonomous categories which model both  linear logic and (some version of) Geometry of Interaction. Apparently, such a category must be partially traced, with the trace somewhat nicely interacting with the $*$-autonomous structure, and the corresponding embedding into a compact category must be consistent with this structure as well.

And indeed, we define for a Mix-category a certain partial trace-like operation on morphisms, which we call  {\it mixed trace}. We show that any structure preserving embedding into a compact category induces a mixed trace, pretty much in the same way as a structure preserving embedding of a monoidal category into a compact one induces a partial trace in the sense of \cite{HS}. Next we show that,  given a mixed trace, we can construct a compact envelope ({\it compactification w.r.t mixed trace}), again very much like the case of a monoidal category and partial trace. Our construction is very similar to that in \cite{Bagnol}. Thus we obtain first necessary and sufficient condition for a Mix-category to embed into a compact one: the Mix-category should have a mixed trace.

However, we also find another, in some sense ``more intrinsic'' condition. It comes from the following considerations.

Structure preserving embedding of a Mix-category into a compact one, essentially consists in adding formal inverses to Mix-maps, i.e. constructing the  fraction category. Of course, in general, the functor to the fraction category is not an embedding. And in a case like ours, if it is an embedding, we can point out certain conditions that must be satisfied. (These conditions might be well-known to experts, but the author did not encounter them in literature.)

Typically if we live in a monoidal category $\bf K$ and add inverses to maps $f_1$, $f_2$, then, in the fraction category
we have for all $h, g_1,g_2$ of corresponding types
$$h\circ(\Id\otimes (f_1^{-1}\circ g_1))\circ ((f_2^{-1}\circ g_2)\otimes\Id )=$$
$$=h\circ((f_2^{-1}\circ g_2)\otimes\Id )\circ(\Id\otimes (f_1^{-1}\circ g_1)). $$
 But both the lefthand and the righthand sides of the above equation may be defined already in ${\bf K}$, and in this case they must be equal in ${\bf K}$. This means that, in ${\bf K}$, any commutative diagram of the form
\begin{diagram}
& & B_1\otimes B_2&&\\
& \ldTo^{f_1\otimes\Id}&\dTo^{h}& \rdTo^{\Id\otimes f_2}& \\
A_1\otimes B_2 & & & & B_1\otimes A_2\\
\uTo^{g_1\otimes\Id}&\rdTo& &\ldTo&\uTo_{\Id\otimes g_2} \\
X_1\otimes B_2 & & R& & B_1\otimes X_2\\
\dTo^{\Id\otimes f_2}&\ruTo & &\luTo   & \dTo_{f_1\otimes\Id}\\
X_1\otimes A_2& & & & A_1\otimes X_2\\
\uTo_{\Id\otimes g_2}&  & & &\uTo_{g_1\otimes\Id}\\
 X_1\otimes X_2 && & &                             X_1\otimes X_2
\end{diagram}
must remain commuting when the lower horizontal arrow is filled-in with the identity morphism.

Generalizing the above condition to  the case when there are more than two morphisms $f_1,f_2,\ldots$, we formulate what we call the {\it contractible zig-zag condition} for a Mix-category, which is necessary for an embedding into a compact one to exist. It turns out that this condition, even some weaker form of it, is also sufficient. When it is satisfied, we manage to define a ``free'' or ``minimal'' mixed trace directly from the structure of  the Mix-category. This gives us  a ``free'' compactification.

It remains an open question if the methods of this paper can be applied to the more general case of a $*$-autonomous category without Mix,  or,  even, simply to a monoidal closed one.

\section{Basics}
In our notation for natural morphisms we often omit sub- and superscripts, when they are clear from the context.

We assume that the reader has some familiarity with symmetric monoidal categories, see \cite{MacLane}.

By default,  monoidal structure is denoted as $\otimes$ and is called tensor product. The monoidal unit is denoted as ${\bf 1}$. Monoidal symmetry is denoted $\sigma_{A,B}:A\otimes B\to B\otimes A$ and is called tensor symmetry.

We recall here what is a {\it monoidal functor}, because this notion will be extensively used.
\bd
A monoidal functor $F=(F,m_{A,B},m_{\bf 1})$ between monoidal categories ${\bf K}$ and ${\bf L}$ is a functor $F:{\bf K}\to{\bf L}$ together with natural transformations $m_{A,B}:F(A)\otimes F(B)\to F(A\otimes B)$ and $m_{\bf {1}}:{\bf 1}\to F({\bf 1})$, satisfying certain coherence conditions, see \cite{MacLane}.
\ed

The monoidal functor  $F=(F,m_{A,B},m_{\bf 1})$ is {\it strong} when $m_{A,B}$, $m_{\bf 1}$ are invertible.
If the categories are symmetric monoidal, the functor is {\it symmetric monoidal} when $$m_{B,A}\circ\sigma_{F(A),F(B)}=F(\sigma_{A,B})\circ m_{A,B}.$$

In this paper, embedding of a symmetric monoidal category is a {\it faithful strong symmetric monoidal functor}.

\subsection{*-Autonomous categories}
$*$-Autonomous categories, introduced by Barr, see \cite{Barr}, are monoidal closed categories with involutive duality. There is a number of equivalent definitions,
here we adopt the following.
\bd
$*$-Autonomous category is a symmetric monoidal category ${\bf K}=({\bf K},\otimes,{\bf 1})$ equipped with a second monoidal structure $\wp$ (cotensor product) and a contravariant functor $(.)^\bot$ (duality), together with a natural isomorphism
 $$A^{\bot\bot}\cong A, $$
 and a natural and dinatural bijection
\be\label{autonomus_closed}
\theta_B^{A,C}:{\bf{K}}(A\otimes B,C)\cong{\bf K}(A,C\wp B^\bot).
\ee
\ed

We will use naturality of $\theta$, so let us recall what does it mean explicitly.

Naturality in $C$: for any $\phi:C\to C'$ and $f:A\otimes B\to C$ we have
$$\theta_B(\phi\circ f)=(\phi\wp B^\bot)\circ\theta_B(f).$$

Naturality in $A$: for any $\psi:A'\to A$ and $g:A\to C\wp B^\bot$ we have
$$\theta_{B}^{-1}(g\circ\psi)=\theta_B^{-1}(g)\circ(\psi\otimes B).$$

 Such a category has the special {\it dualizing object} $\bot={\bf 1}^\bot$, and a number of important maps and isomorphisms coming from  bijection (\ref{autonomus_closed}). Among them, we have isomorphisms:
 $$A\cong A\wp\bot,$$
 De Morgan laws
 $$(A\otimes B)^\bot\cong A^\bot\wp B^\bot,\mbox{}(A\wp B)^\bot\cong A^\bot\otimes B^\bot,$$
 and the maps
 $$ \coev_A:{\bf 1}\to A\wp A^\bot,$$
   $$\ev_A:A\otimes A^\bot\to\bot,$$
   respectively {\it coevaluation} and {\it evaluation}.
\bigskip

A {\it strong $*$-autonomous functor} of $*$-autonomous categories ${\bf K}$ and ${\bf L}$ is a strong symmetric monoidal functor $F:{\bf K}\to{\bf L}$ together with the natural isomorphism $n_A:(F(A))^\bot\to F(A^\bot)$, such that the following diagram commutes.
\begin{diagram}
(F(A^\bot))^\bot& \rTo^{n_{A^\bot}}& F(A^{\bot\bot})\\
\dTo^{(n_A)^\bot}& &\dTo_{\cong}\\
(F(A))^{\bot\bot}&\rTo^{\cong}&F(A)
\end{diagram}
Using De Morgan laws, observe that a strong $*$-autonomous functor is also strong symmetric monoidal with respect to cotensor product, i.e. there is a natural isomorphism $l_{A,B}:F(A)\wp F(B)\to F(A\wp B)$, satisfying necessary coherence conditions.
\bigskip

 For simplifying computations, it is highly desirable to have strict equalities $A=A^{\bot\bot}$, rather than just isomorphisms.  Fortunately, for the purposes of this paper we can always assume that this is the case.
    We have the following definition and theorem.

 \bd
A strict $*$-autonomous category is a $*$-autonomous category, whose monoidal associativity and unit isomorphisms as well as the double duality isomorphisms $A\cong A^{\bot\bot}$ are identities:
$$A\otimes(B\otimes C)=(A\otimes B)\otimes C,$$
$$A\otimes{\bf 1}={\bf 1}\otimes A=A,$$
$$A^{\bot\bot}=A.$$
\ed

\bt\cite{CockettHasegawaSeely}
Any $*$-autonomous category is strongly $*$-autonomously equivalent to a strict one.$\Box$
\et

When the category is strict, a particular instance of isomorphism (\ref{autonomus_closed}) is
\be\label{autonomus_closed_neg}
\theta_{B^\bot}^{A,C}:{\bf{K}}(A\otimes B^\bot,C)\cong{\bf K}(A,C\wp B).
\ee
\bigskip

We denote the tensor and cotensor symmetries as
$$\sigma_{A,B}:A\otimes B\to B\otimes A,\mbox{ }\tau_{A,B}:A\wp B\to B\wp A,$$
and in a strict category we have the identities
\be\label{coev_symmetry}
\ev_{A^\bot}=\ev_{A}\circ\sigma_{A^\bot,A},\mbox{ }\coev_{A^\bot}=\tau_{A,A^\bot}\circ\coev_{A}.
\ee
\bigskip

There is also the {\it  weak distributivity map }
$$\delta=\delta_{A,B,C}:A\otimes(B\wp C)\to (A\otimes B)\wp C, $$
connecting tensor and cotensor structures,
\be\label{delta}
\delta_{A,B,C}=\theta_{C^\bot}(A\otimes\theta^{-1}_{C^\bot}(\Id_{B\wp C})).
\ee
 The above definition of $\delta$ implies the following.
 \nb\label{distributivity_diagram} For any $\phi:A\otimes B\to C$ and object $X$ the following diagram commutes. $\Box$
\begin{diagram}
X\otimes A& &\\
\dTo^{X\otimes\theta\phi}& \rdTo^{\theta(X\otimes\phi)}&\\
X\otimes(C\wp B^\bot)&\rTo^{\delta_{X,C,B^\bot}}& (X\otimes C)\wp B^\bot\\
\end{diagram}
 \nbe
\bigskip

 The weak distributivity map has  other versions as well, obtained by combinations with tensor and cotensor symmetries, such as the {\it right  weak distributivity map }
$$\delta^R=\delta^R_{A,B,C}:(A\wp B)\otimes C\to A\wp (B\otimes C), $$
Iterating distributivities, combined with symmetries, we further get  a number of  important maps, such as:
\be\label{times_rule_map}
(\Id\wp\tau)\circ(\delta\wp\Id)\circ\tau\circ\delta^R\circ(\tau\otimes\Id):(A\wp B)\otimes (C\wp D)\to (A\otimes C)\wp B\wp D.
\ee

\bt\label{times_rule_uniqueness}
Any composition of distributivities and symmetries, resulting in a map of the form $(A\wp B)\otimes (C\wp D)\to (A\otimes C)\wp B\wp D$, results in (\ref{times_rule_map}).
\et
{\bf Proof} This follows from  the defining diagrams for {\it symmetric weakly distributive categories} ($*$-autonomous categories form a subclass of those), see \cite{CockettSeelyWDC}. $\Box$
\bigskip

Bijection $\theta$ in (\ref{autonomus_closed}) is, in fact, defined in terms of composition with weak distributivities and evaluations/coevaluations.
\nb\label{theta from composition}
In a $*$-autonomous category, for
$$\phi:A\otimes B\to C$$
we have
$$\theta \phi=(\phi\wp B^\bot)\circ\delta_{A,B,B^\bot}\circ(A\otimes\coev_B).$$
Respectively, for $$\psi:A\to C\wp B^\bot$$
$$\theta^{-1}\psi=(C\wp\ev_{B^\bot})\circ\delta^R_{C,B^\bot,B}\circ(\psi\otimes B).\mbox{ }\Box$$
\nbe

\subsection{Mix-categories}

\bd\cite{Seely2} A Mix-category is a $*$-autonomous category, equipped with the map
$$\mix:\bot\to{\bf 1}, $$
such that the following diagram commutes.
\begin{diagram}
                      &                   &\bot\otimes\bot&                    & \\
                      &\ldTo^{\mix\otimes\Id}&               &\rdTo^{\Id\otimes \mix}& \\
{\bf 1}\otimes \bot\cong\bot&                   &\rTo_{\Id}     &                    &\bot\cong\bot\otimes {\bf 1}\\
\end{diagram}
\ed

On a Mix-category there are natural {\it mixed evaluation} maps
\be
\widetilde\ev_B=\mix\circ\ev:B\otimes B^\bot\to{\bf 1},
\ee
and {\it Mix-maps}
 $$\Mix_{A,B}:A\otimes B\to A\wp B,$$ defined as
\be\label{expression_for_Mix}
\Mix_{A,B}=\theta_{B^\bot}(A\otimes\widetilde\ev_{B}).
\ee

Observe that the map $\mix:\bot=\bot\otimes{\bf 1}\to\bot\wp{\bf 1}={\bf 1}$ is just a particular instance of a Mix-map.

The distributivity and Mix maps interact well.
\nb\label{mix-distribuitivity}
The diagrams below commute.
\begin{diagram}
A\otimes B\otimes C&\rTo^{\Mix}&(A\otimes B)\wp C\\
\dTo^{\Id\otimes\Mix}&\ruTo_{\delta}&\\
A\otimes(B\wp C)& &\\
\end{diagram}
\begin{diagram}
A\otimes (B\wp C)&\rTo^{\Mix}&A\wp B\wp C\\
\dTo^{\delta}&\ruTo_{\Mix\wp C}&\\
(A\otimes B)\wp C& &\\
\end{diagram}
\nbe
{\bf Proof} The first one commutes by Note \ref{distributivity_diagram}, the second one is, modulo tensor symmetry, the dual of the first. $\Box$
\bigskip

In a Mix-category we define the {\it mixed symmetry} $$\widetilde\sigma_{A,B}:A\otimes B\to B\wp A$$  as
$$ \widetilde\sigma=\tau\circ\Mix=\Mix\circ\sigma.$$ It will play an important role in the sequel.
\bigskip

Finally, let us articulate what exactly is an embedding of a Mix-category. It is a {\it faithful strong $*$-autonomous functor} $F$ of Mix-categories, such that
$$l_{A,B}\circ\Mix_{F(A),F(B)}=F(\Mix_{A,B})\circ m_{A,B}, $$
where $m_{A,B}:F(A)\otimes F(B)\to F(A\otimes B)$, $l_{A,B}:F(A)\wp F(B)\to F(A\wp B)$ are the corresponding natural transformations.

\subsection{Compact categories and traces}

The best known and, probably, best understood class of $*$-autonomous categories is that of {\it compact} (also called {\it compact closed}) ones, whose  two monoidal structures are isomorphic.

A canonical example is the category of finite-dimensional vector spaces and linear maps. Basically, the compact closed structure is an abstraction of the monoidal closed structure of this category.

\bd
A compact category is a $*$-autonomous category in which
$$A\otimes B\cong A\wp B$$
for all objects.
\ed

 We can say that a compact category is a Mix-category where Mix-maps are invertible.

An important feature of a compact category is the {\it categorical trace}. For any morphism $\phi $ of the form $$\phi:A\otimes U\to B\otimes U $$ there exists the {\it trace of $\phi$ over $U$}, the morphism $$Tr_U^{A,B}(\phi):A\to B,$$
defined as
\be\label{compact_trace_def}
Tr_U^{A,B}\phi=\theta_{U^\bot}^{-1}(\phi)\circ (A\otimes\Mix^{-1}_{U,U^\bot})\circ(A\otimes\coev_U).
\ee

(Note that, in our notation, we use the subscript rather than the superscript for the traced object $U$. This seems to us more consistent with mathematical practice; in concrete examples, the trace is often defined in terms of integration or summation over the traced object, which appears in the subscript under  the summation or the integration sign. Also, in speech, we say that we trace the morphism $\phi$ {\it over}, and not {\it under} $U$.)

The above-defined operation is natural and dinatural and satisfies a number of conditions, which provide axioms for a categorical trace. The conditions (including naturality and dinaturality) are:

{\bf Naturality} For  $\phi:A\otimes U\to B\otimes U$,  $f:X\to A$, $g:B\to Y$ it holds that
$$g\circ Tr_U \phi\circ f=Tr_U((g\otimes U)\circ  \phi\circ (f\otimes U)).$$

{\bf Dinaturality w.r.t. symmetries} For $\phi:A\otimes U\otimes V\to B\otimes U\otimes V$
 it holds that $$Tr_{U\otimes V}(\phi)=Tr_{V\otimes U}(\sigma_{U,V}\circ\phi\circ\sigma_{V,U}).$$

{\bf Strength}: For  $\phi:A\otimes U\to B\otimes U$  $f:X\to Y$,  it holds that
$$ f\otimes Tr_U\phi=Tr_U(f\otimes \phi).$$

{\bf Vanishing} For $\phi:A\otimes U\otimes V\to B\otimes U\otimes V$ it holds that
$$Tr_{U\otimes V} \phi=Tr_U(Tr_V \phi). $$

{\bf Yanking}
$$ Tr_U\sigma_{U,U}=\Id_U.$$

(Of course, the above conditions imply that trace is dinatural with respect to all morphisms, not just symmetries.)
\bd A trace on a symmetric monoidal category is a natural and dinatural w.r.t to symmetries operation on hom-sets  $$Tr_U^{A,B}:Hom(A\otimes U,B\otimes U)\to Hom(A,B),$$
satisfying Strength, Vanishing and Yanking.
\ed

{\bf Remark} In literature, the Vanishing axiom is usually supplemented with the condition that for $f:A\to B$ it holds that $Tr_{\bf 1}(a^{-1}_B\circ f\circ a_{A} )=f$, where $a_X: X\otimes{\bf 1}\cong X$.

 But this is, in fact redundant and follows from other axioms. First $(a_B^{-1}\circ f\circ a_{A} )=f\otimes\Id_{\bf 1}$. Next $Tr_{\bf 1}\Id_{{\bf 1}\otimes{\bf{1}}}=Tr_{\bf 1}\sigma_{{\bf 1},{\bf 1}}=\Id_{\bf 1}$ by Yanking. Then,  $Tr^{A,B}_{\bf 1}(f\otimes\Id_{\bf 1})=Tr^{A\otimes{\bf 1},B\otimes{\bf 1}}_{\bf 1}(a_B^{-1}\circ f\circ a_{A} )\otimes\Id_{\bf 1}=Tr^{A\otimes{\bf 1},B\otimes{\bf 1}}_{\bf 1}(f\otimes\Id_{\bf 1}\otimes\Id_{\bf 1})= a_B^{-1}\circ (f\otimes\Id_{\bf 1})\circ a_{A}=f$, using naturality and, in the end, Strength.
 \bigskip

It is well known that  existence of trace characterizes compact categories completely, in the sense that any compact category has a trace, and any category with a trace has canonical full  embedding into a compact one \cite{JoyalStreetVerity}.

{\it Partial trace}, introduced in \cite{HS}, is a generalization of the ordinary (``total'') trace satisfying basically the same properties, but not necessarily defined for all morphisms. Typically, any symmetric monoidal subcategory ${\bf K}$ of a compact ${\bf C}$ has a partial trace. It is defined simply by restricting the canonical total trace of the ambient compact ${\bf C}$ to morphisms of ${\bf K}$, whenever the result is also in ${\bf K}$. It has been proven  \cite{MalherbeScottSelinger},\cite{Bagnol} that partially traced categories are precisely symmetric monoidal subcategories of compact (i.e. totally traced) ones.

Our goal is to characterize Mix-categories which are Mix-subcategories of compact ones. And, as we will see, similarly to the case of monoidal subcategories, one of the characteristics is existence of a certain partial trace-like operation that we call {\it mixed trace}

\section{Necessary conditions for compactification}
In this section we are going to  find necessary conditions for a Mix-category to have a structure preserving embedding into a compact one, a {\it compactification}. We know two such conditions. One of them is expressed in terms of a mixed trace, which is going to be defined shortly. The other one will be discussed right now.

\subsection{Contractible zig-zag condition}
Let ${\bf K}$ be a symmetric monoidal category.

%\begin{diagram}
%& & B_1\otimes\ldots\otimes B_n& & &&  &&&B_{\alpha(1)}\otimes\ldots\otimes B_{\alpha(n)}      &&&\\
%& \ldTo^{f_1\otimes\Id}& &\rdTo(3,6) &&& &&&&&&\\
%A_1\otimes B_2\otimes\ldots\otimes B_n & & & & &&   &&&&&&A_{\alpha(1)}\otimes B_{\alpha(2)}\otimes\ldots\otimes B_{\alpha(n)}\\
%\uTo^{g_1\otimes \Id} &\rdDotsto(5,4) & & & &&\\
%X_1\otimes B_2\otimes\ldots\otimes B_n & & & & &&\\
%\dTo& & & & & &\\
%\cdots& & & \rDotsto& &R\\
%\uTo & &&&\ruDotsto(5,4)   \ruTo(3,6)& &\\
%X_1\otimes\ldots\otimes X_{n-1}\otimes B_n& & & & & &\\
%\dTo^{\Id\otimes f_n}& & &   & & &\\
%X_1\otimes\ldots\otimes X_{n-1}\otimes A_n& & & & &&\\
%& \luTo_{\Id\otimes g_n}&  & & &&\\
%&& X_1\otimes\ldots\otimes X_{n}& && &
%\end{diagram}

\bd\label{zig-zag}
{\rm Assume that we have $3$ tuples of objects $$A_1,\ldots,A_n, \mbox{ } B_1,\ldots,B_n, \mbox{ } X_1,\ldots,X_n,$$
and tuples of morphisms $$f_i:B_i\to A_i, g_i:X_i\to A_i, i=1,\ldots n.$$
The pair of tuples $(f_1,\ldots,f_n)$
and $(g_1,\ldots,g_n)$} satisfies contractible zig-zag condition {\rm if the following holds.}

For any object $R$, any permutation $\alpha\in S_n$ and any collection of morphisms
$$\bigotimes_{j<k}A_j\otimes\bigotimes_{i>k-1}B_i\to R,\mbox{ }k=1,\ldots, n+1,$$
$$\bigotimes_{j<k}A_{\alpha(j)}\otimes\bigotimes_{i>k-1}B_\alpha(i)\to R,\mbox{ }k=1,\ldots, n+1,$$
if the following diagram  commutes,
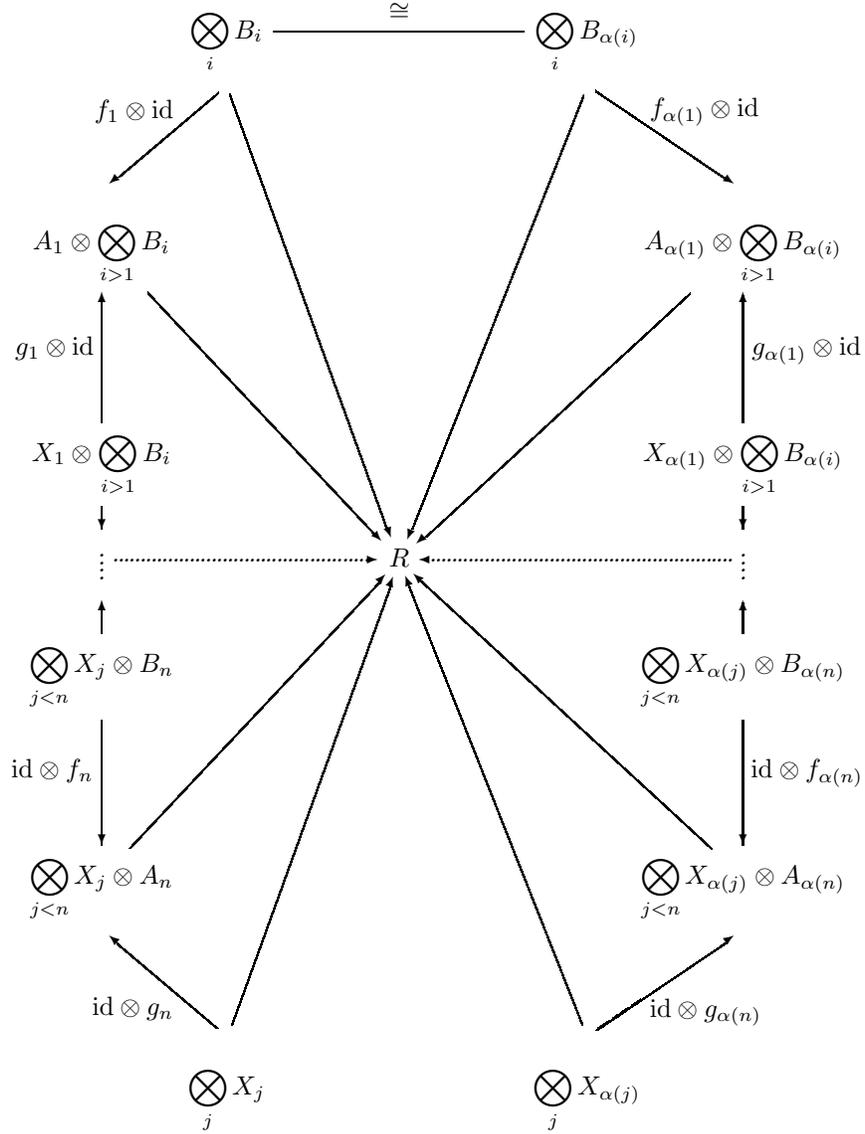
\begin{diagram}
& & \bigotimes_iB_i& & &\rLine^{\cong}&  &&\bigotimes_iB_{\alpha(i)}      &&\\
&&&&&&&&&&\\
& \ldTo^{f_1\otimes\Id}& &\rdTo(3,9) &&& & \ldTo(3,9)&       & \rdTo^{f_{\alpha(1)}\otimes\Id}& \\
&&&&&&&&&&\\
A_1\otimes \bigotimes_{i>1}B_i & & & & &&   &&&&A_{\alpha(1)}\otimes \bigotimes_{i>1}B_{{\alpha(i)}}\\
& \rdTo(5,6)&&&&&&&&\ldTo(5,6)&\\
\uTo^{g_1\otimes \Id} & & & & &&     &&&&\uTo_{g_{\alpha(1)}\otimes \Id}\\
&&&&&&&&&&\\
X_1\otimes \bigotimes_{i>1}B_i & & & & &&    &&&& X_{\alpha(1)}\otimes \bigotimes_{i>1}B_{{\alpha(i)}}\\
\dTo& & & & & &          &&&&\dTo\\
\uDots& & & \rDotsto& &R      &&\lDotsto&&&\uDots\\
\uTo & &&&\ruTo(5,6)\ruTo(3,9) &&\luTo(3,9)\luTo(5,6)         &&&&\uTo\\
\bigotimes_{j<n}X_j\otimes B_n& & & & & &          &&&&\bigotimes_{j<n}X_{\alpha(j)}\otimes B_{\alpha(n)} \\
&&&&&&&&&&\\
\dTo^{\Id\otimes f_n}& & &   & & &                  &&&&\dTo_{\Id\otimes f_{\alpha(n)}}\\
&&&&&&&&&&\\
\bigotimes_{j<n}X_j\otimes A_n& & & & &&            &&&&\bigotimes_{j<n}X_{\alpha(j)}\otimes A_{\alpha(n)}\\
&&&&&&&&&&\\
& \luTo_{\Id\otimes g_n}&  & & &&                     &&&\ruTo_{\Id\otimes g_{\alpha(n)}}& \\
&&&&&&&&&&\\
 &&\bigotimes_{j}X_j && & &                             &&\bigotimes_{j}X_{\alpha(j)}
\end{diagram}
then it remains commuting when the lower horizontal arrow is filled-in with the corresponding tensor symmetry, as below.
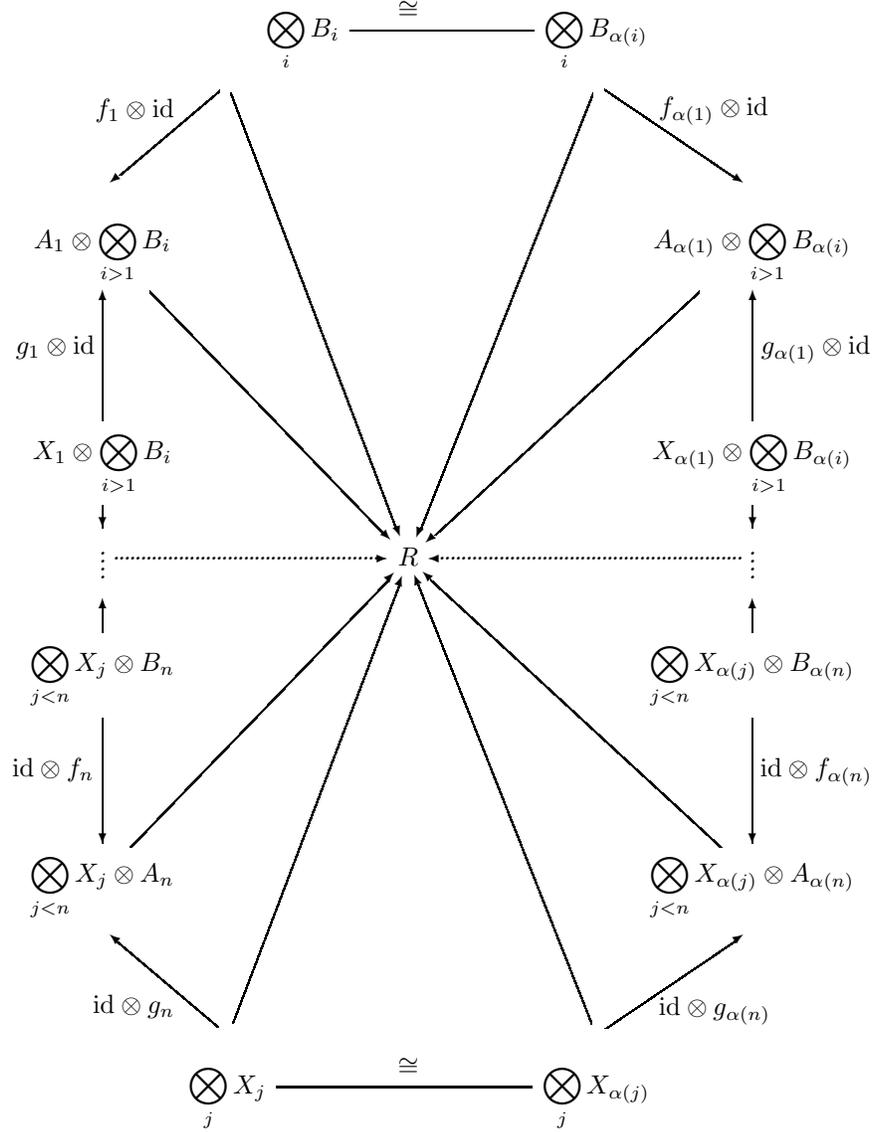
\begin{diagram}
& & &\bigotimes_iB_i& & \rLine^{\cong}&  &&\bigotimes_iB_{\alpha(i)}      &&\\
&&&&&&&&&&\\
& \ldTo^{f_1\otimes\Id}& &\rdTo(3,9) &&& & \ldTo(3,9)&       & \rdTo^{f_{\alpha(1)}\otimes\Id}& \\
&&&&&&&&&&\\
A_1\otimes \bigotimes_{i>1}B_i & & & & &&   &&&&A_{\alpha(1)}\otimes \bigotimes_{i>1}B_{{\alpha(i)}}\\
& \rdTo(5,6)&&&&&&&&\ldTo(5,6)&\\
\uTo^{g_1\otimes \Id} & & & & &&     &&&&\uTo_{g_{\alpha(1)}\otimes \Id}\\
&&&&&&&&&&\\
X_1\otimes \bigotimes_{i>1}B_i & & & & &&    &&&& X_{\alpha(1)}\otimes \bigotimes_{i>1}B_{{\alpha(i)}}\\
\dTo& & & & & &          &&&&\dTo\\
\uDots& & & \rDotsto& &R      &&\lDotsto&&&\uDots\\
\uTo & &&&\ruTo(5,6)\ruTo(3,9) &&\luTo(3,9)\luTo(5,6)         &&&&\uTo\\
\bigotimes_{j<n}X_j\otimes B_n& & & & & &          &&&&\bigotimes_{j<n}X_{\alpha(j)}\otimes B_{\alpha(n)} \\
&&&&&&&&&&\\
\dTo^{\Id\otimes f_n}& & &   & & &                  &&&&\dTo_{\Id\otimes f_{\alpha(n)}}\\
&&&&&&&&&&\\
\bigotimes_{j<n}X_j\otimes A_n& & & & &&            &&&&\bigotimes_{j<n}X_{\alpha(j)}\otimes A_{\alpha(n)}\\
&&&&&&&&&&\\
& \luTo_{\Id\otimes g_n}&  & & &&                     &&&\ruTo_{\Id\otimes g_{\alpha(n)}}& \\
&&&&&&&&&&\\
 &&\bigotimes_{j}X_j && & \rLine^{\cong}&                             &&\bigotimes_{j}X_{\alpha(j)}
\end{diagram}
\ed

\nb If $f_1,\ldots,f_n$ are morphisms in a symmetric monoidal category ${\bf K}$ and $F$ is a faithful strong symmetric monoidal functor from ${\bf K}$ into a symmetric monoidal category ${\bf K}'$, where images of $f_1,\ldots,f_n$ are invertible, then for any collection $(g_1,\ldots,g_n)$ of ${\bf K}$-morphisms the pair $(f_1,\ldots,f_n)$ and $(g_1,\ldots,g_n)$ satisfies contractible zig-zag condition.
\nbe
{\bf Proof} Identifying all morphisms with their images under $F$, we compute morphisms of the first diagram in Definition \ref{zig-zag} in ${\bf K}'$. The leftmost zig-zag path from $\bigotimes_{j}X_j$ to $\bigotimes_iB_i$ reads as the morphism,
$(f_1^{-1}\circ g_1)\otimes\ldots\otimes(f_n^{-1}\circ g_n)$, and the rightmost zig-zag path from $\bigotimes_{j}X_{\alpha(j)}$ to $\bigotimes_iB_{\alpha(i)}$  as $(f_{\alpha(1)}^{-1}\circ g_{\alpha(1)})\otimes\ldots\otimes(f_{\alpha(n)}^{-1}\circ g_{\alpha(n)})$. The statement follows. $\Box$
\bigskip

\bc
If a Mix-category $\bf K$ embeds as a Mix-category into a compact one, then, for any tuples of objects $A_1,\ldots,A_n$, $B_1,\ldots,B_n$, $X_1,\ldots,X_n$ and morphisms $g_i:X_i\to A_i\wp B_i$, the pair $(\Mix_{A_1,B_1},\ldots,\Mix_{A_n,B_n})$ and $(g_1,\ldots, g_n)$ satisfies contractible zig-zag condition. $\Box$
\ec

This gives us first necessary condition for existence of compactification of a Mix-category, and we are going to show that it is sufficient. In fact, we will need it in a considerably weaker form.
\bc\label{contractible mix}
If a Mix-category $\bf K$ embeds as a Mix-category into a compact one, then, for any tuple of objects $A_1,\ldots,A_n$,  the pair $(\Mix_{A_1,A^\bot_1},\ldots,\Mix_{A_n,A^\bot_n})$ and $(\coev_{A_1},\ldots, \coev_{A_n})$ satisfies contractible zig-zag condition. $\Box$
\ec

We will call categories satisfying the conditions of Corollary \ref{contractible mix} {\it contractible zig-zag Mix-categories}.

In the next subsection we find another necessary condition, expressed in terms of existence of a certain partial trace.

\subsection{Mixed trace}
A structure preserving embedding of a Mix-category into a compact one equips the former  not only with a partial trace in the sense of \cite{HS}, but with a more general trace-like operation that we call {\it mixed trace}.
In order to describe it we introduce some notation and terminology.

For objects $A,B\in {\bf K}$ we define a {\it loop} $p:A\looparrowright B$ as a tuple $$p=(\phi;U_1,\ldots,U_k),$$ where $k\in{\bf N}$,  $U_i$, $i=1,\ldots k$, the {\it hidden part}, are   objects of ${\bf K}$, and $\phi$, the {\it carrier}, is a ${\bf K}$-map $$\phi:A\otimes U_1\otimes\ldots\otimes U_k\to B\wp U_1\wp\ldots\wp U_k.$$ The number $k$ in the above definition can equal $0$, in which case the hidden part is empty, and the corresponding morphism is just a ${\bf K}$-morphism from $A$ to $B$. Thus, a ${\bf K}$-morphism  is identified as a loop with the empty hidden part.

Clearly a structure preserving embedding into a compact category allows tracing loops over their hidden parts (with the convention that  tracing over the empty  tuple does nothing)

Let us use the following vector notation. We denote a tuple of objects as
$$\vec{U}=(U_1,\ldots, U_n),$$
with the conventions that
$$A\wp\vec{U}=A\wp U_1\wp\ldots\wp U_n,\mbox{ }A\otimes\vec{U}=A\otimes U_1\otimes\ldots\otimes U_n,$$
$$\Vec{U}^\bot=(U_1^\bot,\ldots,U_n^\bot),$$
$$\tau_{A,\vec{U}}=\tau_{A,U_1\wp\ldots\wp U_n},\mbox{ }\sigma_{A,\vec{U}}=\sigma_{A,U_1\otimes\ldots\otimes U_n},$$
and $\Mix_{A,\vec{U}}$ is the obvious iteration of Mix-maps
$$\Mix_{A,\vec{U}}:A\otimes U_1\otimes\ldots\otimes U_n\to A\wp U_1\wp\ldots\wp U_n.$$
Also, if $\vec{V}=(V_1,\ldots,V_k)$, then
$$\vec{V}\wp\vec{U}=V_1\wp\ldots\wp V_k\wp U_1\wp\ldots\wp U_n,\mbox{ }\vec{V}\otimes\vec{U}=V_1\otimes\ldots\otimes V_k\otimes U_1\otimes\ldots\otimes U_n,$$
and so on.
If $\alpha\in S_n$ is a permutation, then $\alpha\vec{U}=(U_{\alpha(1)},\ldots,U_{\alpha(n)})$.

Now, If $F$ is the embedding, with the corresponding natural transformations $m_{A,B}:FA\otimes FB\to F(A\otimes B)$, $l_{A,B}:FA\wp FB\to F(A\wp B)$, and $$p=(\phi;\vec{U}):A\looparrowright B$$ is a loop, then the {\it mixed trace} of $p$ over $\vec{U}$, is the map, which we, abusing notation, still denote as $Tr(p):A\to B$,  defined by the equation
\be\label{traces agree}
F(Tr (p))=Tr_{F(U_1)\otimes\ldots F(U_n)}^{F(A),F(B)}(\Mix_{F(B),F(\vec{U})}^{-1}\circ  l^{-1}_{B,\vec{U}}
\circ  F(\phi)\circ m_{A,\vec{U}}),
\ee
whenever it has a solution.

The above defined partial operation enjoys certain conditions, which we read from the conditions for trace. In order to write them concisely we introduce certain operations on loops.

{\bf Composition with a morphism} For the loop $p=(\phi;\vec{U}):A\looparrowright B$, and morphisms $f:X\to A$, $g:B\to Y$  the compositions $$g\circ p:A\looparrowright Y, \mbox{} p\circ f:X\looparrowright B$$ are defined by
$g\circ p=((g\wp \vec{U})\circ \phi;\vec{U})$, $p\circ f=(\phi\circ(f\wp \vec{U});\vec{U})$.

{\bf Multiplication by a morphism}: For a loop $p=(\phi;\vec{U}):A\looparrowright B$ and morphism $f: C\to D$, the loop $$f\otimes p:C\otimes A\looparrowright D\otimes B$$ is defined by $f\otimes p=(\delta\circ (f\otimes\phi);\vec{U})$.

 {\bf Hidden symmetry}  For the loop $p=(\phi;U_1,\ldots, U_k):A\looparrowright B$ and a permutation $\alpha\in S_n$ we define the loop $$\alpha p:A\looparrowright B$$ by $\alpha p=((\Id\wp\tau_\alpha)\circ\phi\circ(\Id\otimes\sigma_{\alpha^{-1}});\alpha\vec{U})$, where $\sigma_\alpha$, $\tau_\alpha$ are the obvious tensor and cotensor symmetry.

{\bf Dual} of the loop $p=(\phi;\vec{U}):A\looparrowright B$ is the loop $p^\bot=(\phi^\bot;\vec{U}^\bot):B^\bot\looparrowright A^\bot$.

\bd A mixed trace is a partial operation on loops, mapping a loop  $p=(\phi,\vec{U}):A\looparrowright B$ to a morphism $Tr_{\vec{U}}^{A,B}p:A\to B$, satisfying the following conditions:

{\bf Naturality} For  $p=(\phi;\vec{U}):A\looparrowright B$,  $f:X\to A$, $g:B\to Y$ it holds that
$$g\circ Tr( p)\circ f=Tr(g\circ  p\circ f),$$
whenever the lefthand side is defined.

{\bf Dinaturality w.r.t. symmetries} For the loop $p$ with the hidden part $\vec{U}=(U_1,\ldots,U_n)$ and a permutation $\alpha\in S_n$ it holds that $$Tr(p)=Tr(\alpha p),$$
whenever any side of the equation is defined.

{\bf Strength}: For  $p:A\looparrowright  B$,  $\phi:X\to Y$,  it holds that
$$ \phi\otimes Tr(p)=Tr(\phi\otimes p),$$
whenever the lefthand side is defined.

{\bf Vanishing} For $p=(\phi;\vec{U},\vec{V}):A\looparrowright  B$, $q=(\phi;\vec{V}):A\otimes\vec{U}\looparrowright  B\wp\vec{U}$, if $Tr( q)$ exists, then
$$Tr( p)=Tr(Tr (q)), $$
whenever any side of the equation is defined.

{\bf Adjointability} $$Tr(p)^\bot=Tr(p^\bot),$$
whenever any side of the equation is defined.

{\bf Yanking} For $p=(\widetilde\sigma_{A,A}, A):A\looparrowright A$ it holds that
$$ Tr( p)=\Id_A.$$
\ed

\bl\label{trace_necessary} If a Mix-category has a structure preserving embedding into a compact one, then it admits a mixed trace. $\Box$
\el

Our next goal is to show that the converse statement also holds: if a Mix-category ${\bf K}$ has a mixed trace, then it embeds into a compact one. This is proven by constructing a compact envelope of ${\bf K}$, whose objects are objects of ${\bf K}$, and morphisms are loops of ${\bf K}$, quotiented by a certain equivalence relation. Basically, it is the smallest equivalence, compatible with the Mix-category structure and  the given mixed trace. The construction  is very similar to the analogous construction in \cite{Bagnol}.

\section{Compactification from mixed trace}
In this section we show that having a mixed trace is also a sufficient condition for a Mix-category to have an embedding into a compact one.

We consider a Mix-category ${\bf K}$ and assume that a mixed trace is defined on it. We are going to  construct a larger category, the { \it compactification of ${\bf K}$ with respect to the given mixed trace}, which is compact and contains ${\bf K}$ as a subcategory. Morphisms in this new category are constructed from loops in ${\bf K}$.

\subsection{Category of loops}
We want to organize loops on $\bf K$ into a Mix-category, which later will be quotiented by a certain equivalence relation and become compact.

For that purpose we define a number of operations on loops

{\bf Tensor product} For loops $p=(\phi;\vec{U}):A\looparrowright B$, $q=(\psi,\vec{V}): C\looparrowright D$, their {\it tensor product} $$p\otimes q:A\otimes C\looparrowright B\otimes D$$ is the loop with the hidden part $(\vec{U},\vec{V})$  and the carrier $\widehat{\phi\otimes \psi}$ defined by the composition
\begin{diagram}
A\otimes \vec{U}\otimes B\otimes \vec{V} &\rTo^{\phi\otimes\psi} &(A\wp \vec{U})\otimes( B\wp \vec{V})\\
\uTo_{\Id\otimes\sigma\otimes\Id}& & \dTo^{}\\
A\otimes B\otimes \vec{U}\otimes \vec{V}&\rTo_{\widehat{\phi\otimes\psi}} &( A\otimes B)\wp \vec{U}\wp \vec{V}, \\
\end{diagram}
where the right vertical arrow is obtained as a composition of symmetries and distributivity maps (there is no ambiguity in its definition by Theorem \ref{times_rule_uniqueness}).

Note that it follows from the same theorem that tensor product of loops is associative (remember that we work in a strict category)

Note that this operation extends tensor multiplication by morphisms, defined it the preceding section.

{\bf Cotensor product} is defined by tensor and duality (duality was defined in the preceding section).  For loops $p=(\phi;\vec{U}):A\looparrowright B$, $q=(\psi,\vec{V}): C\looparrowright D$, their {\it cotensor product} $$p\wp q:A\wp C\looparrowright B\wp D$$ is the loop $$p\wp q=(p^\bot\otimes q^\bot)^\bot.$$

Note that for loops with empty hidden parts,  i.e. usual ${\bf K}$-morphisms the above are the usual operations on morphisms.

{\bf Hiding} For the loop $p=(\phi;\vec{U}):A\otimes V\looparrowright B\wp V$ we define the new loop
$$\Hid^{A,B}_V(p)=(\phi;V,\vec{U}):A\looparrowright B.$$

{\bf Hidden trace} This  is  a partially defined operation.
 For the loop $$p=(\phi;\vec{U},\vec{V}):A\looparrowright B$$ let $q=(\phi;\vec{V}):A\otimes\vec{U}\looparrowright B\wp\vec{U}$. The hidden trace $Tr_{V}(p)$ over $V$ is defined as the mixed trace followed by hiding
  $$Tr_{V}p=\Hid^{A,B}_UTr(q).$$

{\bf Composition} For the loops $p=(\phi;\vec{U}):A\looparrowright B$, $q=(\psi;\vec{V}):B\looparrowright C$, their composition  is the loop $q\circ p:A\looparrowright C$, with the hidden part $\vec{U},\vec{V}$ and the carrier $\xi$ defined by the diagram

%\begin{diagram}
%& &(V_1\otimes\ldots\otimes V_l\otimes B)\wp U_1\wp\ldots\wp U_k & &\\
%&\ruTo & &\rdTo_{\sigma\wp\Id}& \\
%V_1\otimes\ldots\otimes V_l\otimes(B\wp U_1\wp\ldots\wp U_k)&& &&(B\otimes V_1\otimes\ldots \otimes V_l)\wp U_1\wp\ldots\wp U_k\\
%\uTo_{\Id\otimes\phi}&      & &                              &\dTo^{\psi\wp\Id}\\
%V_1\otimes\ldots\otimes V_l\otimes A\otimes U_1\otimes\ldots\otimes U_k & & &  &C\wp V_1\wp\ldots\wp V_l\wp U_1\wp\ldots\wp U_k\\
%\uTo_{\sigma}&       & &                            &\dTo^{\tau}\\
%A\otimes B\otimes U_1\otimes\ldots U_k\otimes V_1\otimes\ldots\otimes V_l & &\rTo^{\xi} & &C\wp U_1\wp\ldots U_k\wp V_1\wp\ldots\wp V_l. \\
%\end{diagram}
%

\begin{diagram}
(B\wp \vec{U})\otimes \vec{V}  & \rTo &(B\otimes \vec{V})\wp \vec{U}\\
&                                 &\dTo_{\psi\wp\Id}\\
\uTo^{\phi\otimes\Id}& & C\wp \vec{V}\wp \vec{U}\\
&                                 &\dTo_{\Id\wp\tau}\\
A\otimes \vec{U}\otimes \vec{V}&  \rTo_{\xi} & C\wp \vec{U}\wp \vec{V}, \\
\end{diagram}
where the upper horizontal arrow is obtained as a composition of symmetry and distributivity maps (there is no ambiguity in its definition by Theorem \ref{times_rule_uniqueness}).

Note that it follows from the same Theorem \ref{times_rule_uniqueness} and naturality of symmetries and distributivities that composition of loops is associative.

Note that this composition extends composition with morphisms, defined in the preceding section. In particular the assignment $f\mapsto (f;)$, sending morphisms to loops, is functorial.
\bigskip

It follows that the category, whose objects are objects of $\bf K$, and morphisms are loops, is well-defined, has monoidal structure and duality. The underlying category ${\bf K}$ embeds into the category of loops as a symmetric monoidal category. In fact, the category of loops is $*$-autonomous and Mix, and the embedding of ${\bf K}$ preserves the structure.

Indeed, by Note \ref{theta from composition}, defining isomorphism (\ref{autonomus_closed}) in a $*$-autonomous category is realized by composition with the natural weak distributivity and evaluation or coevaluation maps, and the category of loops inherits these maps from ${\bf K}$. In the same way it inherits Mix-maps.

\subsection{Congruence and loop operations}
We are going  to define a certain equivalence relation on loops and see how it interacts with  loop operations.

For any two objects $A,B$ and loops  we define the {\it one-step congruence}  relation $\smile$ on loops $A\looparrowright B$ by

(i) a loop is one-step congruent to its hidden trace;

(ii) loops, related by a hidden symmetry are one-step congruent.

{\it Loop congruence} is the  equivalence relation, generated by the  one-step congruence.

\nb
If $p_1\smile p_2$ then for any morphism $\phi$ it holds that $\phi\otimes p_1\smile\phi\otimes p_2$.
\nbe
{\bf Proof} If $p_1$ and $p_2$ are related by a hidden symmetry, the claim follows from naturality of the weak distributivity map. Otherwise it follows from the Strength property of  mixed trace. $\Box$

\nb
For loops $p_1,p_2:A\looparrowright B$ and morphisms $f:B\to Y$, $g:X\to A$, if $p_1\smile p_2$ then  $f\circ p_1\circ g\smile f\circ p_2\circ g$.
\nbe
{\bf Proof} If the loops are related by a hidden symmetry, the claim is obvious. otherwise it follows from naturality of mixed trace. $\Box$

The two notes above imply
\bl Loop congruence is  preserved by compositions and tensor products with morphisms. $\Box$
\el

The following is obvious.
\nb Loop congruence is preserved by hiding. $\Box$
\nbe

Now,  tensor product of loops $p$ and $q$ is, in fact, nothing else than the tensor product of the carrier of $p$  (considered as a loop with the empty hidden part, i.e. an ordinary morphism) with $q$, followed by composition with a tensor symmetry on the left and a weak distributivity  on the right and then by hiding. Since all these operations preserve loop congruence, it follows that tensor product with a loop preserves loop congruence.
\bl
Tensor product of loops preserves loop congruence. $\Box$
\el

Hidden trace preserves duality by the Adjointability condition. Hidden symmetry preserves duality as well, i.e., for a loop $p$ and any permutation $\alpha$ on its hidden part, we immediately see that
 $$(\alpha p)^\bot=\alpha^{-1} p^\bot.$$

 This, together with the preceding Lemma yields us the following.
\bl
Duality and cotensor product of loops preserve loop congruence. $\Box$
\el

Finally, for loops $p=(\phi;\vec{U}):A\looparrowright B$ and $q=(\psi;\vec{V}):B\looparrowright C$, their composition is obtained from the loop $\psi\circ\sigma_{\vec{V},B}\circ(\Id_{\vec{V}}\otimes p)$, by composing it with $\sigma_{A,\vec{V}}$ on the left, then hiding $\vec{V}$ and applying hidden symmetry (i.e. permuting $\vec{U}$ and $\vec{V}$). Again, all operations involved preserve congruence, so composition with a loop preserves congruence as well.

\bl
Loop congruence is preserved by composition of loops. $\Box$
\el

\subsection{Compactification}
From the above  it follows that we can organise a well-defined Mix-category ${\bf C_{Tr}(K)}$ with the same objects as in ${\bf K}$ and morphisms being equivalence classes of loops with respect to the loop congruence. We call this category the {\it compactification} of ${\bf K}$ {\it with respect to the mixed trace}.

The functor $C_{Tr}:{\bf K}\to{\bf C_{Tr}(K)}$ is defined which is identity on objects and sends a  morphism
$\phi$ to the equivalence class of $(\phi;)$.

We now show that this functor  is faithful.
This follows from the two straightforward lemmas below.

\bl
If $(\phi;)\smile p$, then  $\phi=Tr(p)$. $\Box$
\el

\bl
If $Tr(p)=\phi$ and $q\smile p$, then $Tr(q)=\phi$, (here, $\phi$ is a ${\bf K}$-morphism).
\el
{\bf Proof} If $p$ and $q$ are related by a hidden symmetry, this follows from dinaturality of mixed trace w.r.t. symmetries. Otherwise it follows from Vanishing. $\Box$
\bigskip

Finally, let us show that in the compactification ${\bf C_{Tr}(K)}$ the two monoidal structures become isomorphic, hence ${\bf C_{Tr}(K)}$ is compact.

\bl
In ${\bf C_{Tr}(K)}$ the Mix-map has inverse.
\el
{\bf Proof} The inverse of $\Mix:A\otimes B\to A\wp B$ is the  loop
$$\rm{coMix}_{A,B}=(\widetilde\sigma_{A\wp B,A\otimes B}; A,B):A\wp B\looparrowright A\otimes B.$$
It is  sufficient to note that $$\rm{coMix}_{A,B}\circ \Mix_{A,B}=\Hid_A(\widetilde\sigma_{A,A})\otimes \Hid_B(\widetilde\sigma_{B,B}).$$ This is established by a routine diagram chasing, using repeated iteration of Note \ref{mix-distribuitivity} together with Theorem \ref{times_rule_uniqueness}.
But by Yanking, $$\Hid_X(\widetilde\sigma_{ X,X})\smile\Id_X,$$ and, since tensor product of loops preserves congruence, we conclude that in ${\bf C_{Tr}(K)}$ it holds that $\rm{coMix}\circ \Mix=\Id$.
Then by duality $\Mix\circ\rm{coMix}=\Id$ as well. $\Box$

Combining the above with Lemma \ref{trace_necessary} we get the following

\bt\label{compactification from trace} A Mix-category embeds as a Mix-category into a compact one, if and only if it admits a mixed trace. $\Box$
\et

{\bf Remark} The constructed compactification is free with respect to the given mixed trace, i.e. any Mix-categories functor $F$   from ${\bf K}$ to  a compact ${\bf C}$  that agrees with this mixed trace in the sense of equation (\ref{traces agree}) factors through $C_{Tr}$. We define the functor
$F':{\bf C(K)}\to{\bf K}$  sending the loop $p=(\phi;\vec{U}):A\looparrowright B$ to $$F'(p)=Tr_{F(U_1)\otimes\ldots F(U_n)}^{F(A),F(B)}(\Mix_{F(B),F(\vec{U})}^{-1}\circ  l^{-1}_{B,\vec{U}}
\circ  F(\phi)\circ m_{A,\vec{U}}),$$ (here $m:F(X)\otimes F(Y)\to F(X\otimes Y)$, $l:F(X)\wp F(Y)\to F(X\wp Y)$ are the corresponding natural transformations), and then
$F=F'\circ C_{Tr}$.
\bigskip

\section{Mixed trace from contractible zig-zag condition}
We have shown that having a mixed trace is both necessary and sufficient condition for a Mix-category to admit a structure preserving embedding into a compact one. Now we will show that  contractible zig-zag condition is sufficient as well. Using this condition we will define an intrinsic ``free'' mixed trace on a Mix-category, which is equivalent to having a compactification.

Let ${\bf K}$ be a contractible zig-zag Mix-category.

Let $$p=(\phi;\vec{U}):A\looparrowright B,\mbox{ }\vec{U}=(U_1,\ldots,U_n),$$ be a loop in ${\bf K}$.

We first define the {\it provisional trace} of $p$.

Consider the morphism $$ \phi': \bigotimes_i(U_i\otimes U_i^\bot)\to B\wp A^\bot,$$ obtained from $\phi$ by iterated bijection (\ref{autonomus_closed}) and tensor symmetries.

 \bd Provisional trace $\widehat{Tr}(p):A\to B$ of the loop $p:A\looparrowright B$ exists, if there exists a morphism $$\psi:{\bf 1}\to B\wp A^\bot$$ and $n$ morphisms
 $$(U_i\wp U_i^\bot)\otimes\bigotimes_{j>i}(U_j\otimes U_j^\bot)\to B\wp A^\bot,\mbox{ }i=1,\ldots, n, $$
 which, together with $\phi'$ fit into the following commutative diagram.

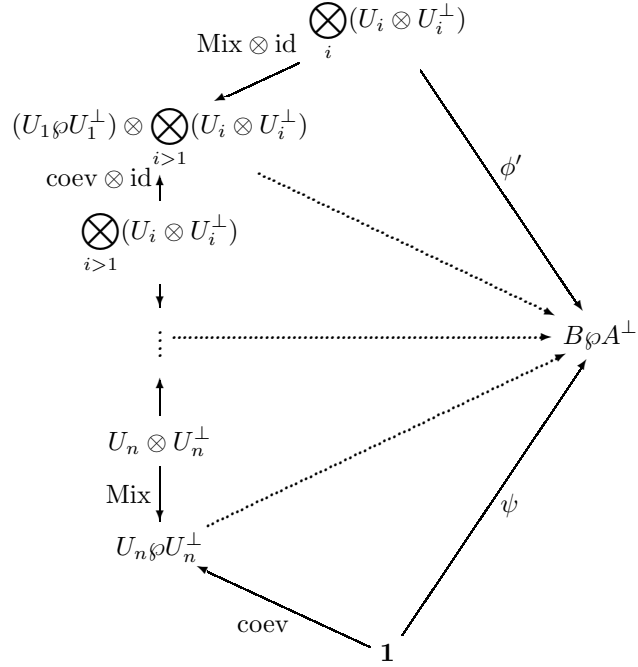
\begin{diagram}\label{trace_def}
& & \bigotimes_i(U_i\otimes U_i^\bot)& & &&\\
& \ldTo^{\Mix\otimes\Id}& &\rdTo(3,6)^{\phi'} &&&\\
(U_{1}\wp U_{1}^\bot)\otimes\bigotimes_{i>1}(U_i\otimes U_i^\bot) & & & & &&\\
\uTo^{\coev\otimes \Id} &\rdDotsto(5,4) & & & &&\\
\bigotimes_{i>1}(U_i\otimes U_i^\bot) & & & & &&\\
\dTo& & & & & &\\
\uDots& & & \rDotsto& &B\wp A^\bot\\
\uTo & &&&\ruDotsto(5,4)   \ruTo(3,6)_{\psi}& &\\
U_{n}\otimes U_{n}^\bot& & & & & &\\
\dTo^{\Mix}& & &   & & &\\
U_{n}\wp U_{n}^\bot& & & & &&\\
& \luTo_{\coev}&  & & &&\\
&& {\bf 1}& && &
\end{diagram}
In which case  $$\widehat{Tr}(p)=\theta_A^{-1}(\psi).$$
\ed

\bd Mixed trace $Tr(p):A\to B$ of the loop $p:A\looparrowright B$ exists, if for some permutation $\alpha \in S_n$ the provisional trace $\widehat{Tr}(\alpha p)$ exists. In which case $$Tr(p)=\widehat{Tr}(\alpha p)$$ for this $\alpha$.
\ed

The contractible zig-zag condition guarantees that the mixed trace, whenever defined, is defined unambiguously.
\bigskip

Obviously, we can give an alternative definition of provisional trace.
\nb\label{trace_def_dual}
For $p:A\looparrowright B$ the provisional trace $\widehat{Tr}(p)$ can be equivalently defined using the following diagram,
\begin{diagram}
& & \bigwp_i(U_i^\bot\wp U_i)& & &&\\
& \ruTo^{\Mix\wp\Id}& &\luTo(3,6)^{(\phi')^\bot} &&&\\
( U_{1}^\bot\otimes U_{1})\wp\bigwp_{i>1}(U_i^\bot\wp U_i) & & & & &&\\
\dTo^{\ev\wp \Id} &\luDotsto(5,4) & & & &&\\
\bigwp_{i>1}(U_i^\bot\wp U_i) & & & & &&\\
\uTo& & & & & &\\
\cdots& & & \lDotsto& &B^\bot\otimes A\\
\dTo & &&&\ldDotsto(5,4)   \ldTo(3,6)_{\psi^\bot}& &\\
U_n^\bot\wp U_n& & & & & &\\
\uTo^{\Mix}& & &   & & &\\
U_n^\bot\otimes U_n& & & & &&\\
& \rdTo_{\ev}&  & & &&\\
&& {\bf \bot}& && &
\end{diagram}
by $$\widehat{Tr}(p)=(\theta_A(\psi^\bot))^\bot.\Box$$
\nbe
\bigskip

We need to check that the above defined operation is indeed a mixed trace.

Indeed, naturality is immediate from definition. Dinaturality w.r.t. symmetries and Vanishing follow from the contractible zig-zag condition. Strength follows from naturality of the weak distributivity map. Adjointability follows from above Note \ref{trace_def_dual}. It remains to establish Yanking.

\bl\label{mixed_yanking}
The above defined mixed trace satisfies Yanking axiom.
\el
{\bf Proof} Let $p=(\widetilde\sigma_{U,U};U)$.

Let
\be
\sigma^{(1)}=\theta_{U^\bot}^{-1}(\widetilde\sigma_{U,U}):U\otimes U\otimes U^\bot\to U,
\ee

\be\label{sigma^2}
\sigma^{(2)}=\sigma^{(1)}\circ\sigma_{U\otimes U^\bot,U}:U\otimes U^\bot\otimes U\to U,
\ee

\be
\sigma^{(3)}=\theta_U(\sigma^{(2)}):U\otimes U^\bot\to U\wp U^\bot.
\ee

The last map $\sigma^{(3)}$ plays the role of $\phi'$ in Definition \ref{trace_def} of provisional trace (and $\widetilde\sigma$ plays the role of $\phi$).

Now,
$$\sigma^{(1)}=(U\otimes\widetilde\ev_U)\circ(\sigma_{U,U}\otimes U^\bot)$$
by naturality of $\theta$,
$$\sigma^{(2)}=(U\otimes\widetilde\ev_U)\circ(\sigma_{U,U}\otimes U^\bot)\circ\sigma_{U\otimes U^\bot,U}=$$
$$=(U\otimes\widetilde\ev_U)\circ(U\otimes\sigma_{U^\bot,U})=U\otimes\widetilde\ev_{U^\bot}$$
by  (\ref{coev_symmetry}), and
 $$\sigma^{(3)}=\Mix_{U,U^\bot},$$
 from defining equation (\ref{expression_for_Mix}).

 Thus we get the commutative diagram
 \begin{diagram}
U\otimes U^\bot& & &\\
\dTo^{\Mix}&\rdTo^{\sigma^{(3)}}& &\\
(U\wp U^\bot)&\rDotsto^{\Id}& U\wp U^\bot\\
\uTo^{\coev}&\ruTo_{\coev}&&\\
{\bf 1},& & &\\
\end{diagram}
and by Definition \ref{trace_def}
$\widehat{Tr}(p)=\theta^{-1}(\coev)=\Id$. The statement follows. $\Box$
\bigskip

Thus, the above-defined operation is indeed a mixed trace. It follows then from Theorem \ref{compactification from trace} that a Mix-category embeds, as a Mix-category, into a compact one if and only if it is a contractible zig-zag Mix-category.
\bigskip

{\bf Remark} The constructed mixed trace is free in the sense that any embedding into a compact category will always induce a mixed trace extending this one. This is clear from definition (\ref{compact_trace_def}) of trace on a compact category. Consequently, the compactification of a contractible zig-zag Mix-category defined by this trace is free as well, in the sense that any Mix-category structure preserving functor of the given category into a compact one factors through this embedding.
\bigskip

We summarize with the following.

\bt
For a Mix-category ${\bf K}$ the following are equivalent:

$(i)$ ${\bf K}$ embeds as a Mix-category into a compact one;

$(ii)$ in ${\bf K}$, for any $n$-tuple of  mix-maps $\Mix_{A_i,B_i}$, $i=1,\ldots,n$ and an $n$-tuple of arbitrary maps $f_i$ with codomains $A_i\wp B_i$, $i=1,\ldots,n$, this pair of  tuples satisfies contractible zig-zag condition;

$(iii)$ in ${\bf K}$ any pair  $(\Mix_{A_1,A^\bot_1},\ldots,\Mix_{A_n,A^\bot_n})$ and $(\coev_{A_1},\ldots,\coev_{A_n})$  satisfies contractible zig-zag condition;

$(iv)$ ${\bf K}$ has a mixed trace. $\Box$
\et
\bigskip

\section{Some concluding remark}
We have found necessary and sufficient conditions for a Mix-category to have a structure preserving embedding into a compact one and defined mixed trace, a generalized partial trace suitable for this setting. We also constructed free mixed trace and free compactification.

 We did not solve, however, the question when does a $*$-autonomous category have a {\it $*$-autonomous} embedding into a compact one, and what is the corresponding generalized partial trace.

 It is worth noting that one and the same $*$-autonomous category can have different non-isomorphic Mix-structures, and different compactifications, corresponding to them. For example, on the compact category of free finitely generated ${\bf Z}$-modules, where tensor and cotensor product are equal, we can define natural Mix-maps as multiplications by a fixed integer $n$. When $n=0$, there is no compactification at all, otherwise, the free compactification is quite obviously the same category localized away from $n$, i.e. the category of free finitely generated modules over ${\bf Z}[\frac{1}{n}]$. Thus, compactification determined by the Mix- structure is not the same as compactification determined by $*$-autonomous structure.

 It seems that methods of this paper, with slight technical modifications, can be applied to the general $*$-autonomous setting, but this has to be worked out. So far, compactification of a $*$-autonomous category is an open  question for us.

 One can proceed further and see what happens when $*$-autonomous categories are replaced with general monoidal closed ones.

 These topics are left for future work.

\end{document}